# Research on Life Characteristics of Internet Based on Network Motifs

Jinfa Wang, Hai Zhao, Xiao Liu

School of Computer Science and Engineering, Northeastern University, Shenyang, China.

**Abstract**: From biosystem to complex system, the study of life is always an important area. Inspired by hyper-cycle theory about the evolution of non-life system, we study the metabolism, self-replication and mutation behavior in the Internet based on node entity, connection relationship and function subgraph—motif—of network topology. Firstly a framework of complex network evolution is proposed to analyze the birth and death phenomena of Internet topology from January 1998 to August 2013. Then we find the Internet metabolism behavior from angle of node, motif to global topology, i.e. one born node is only added into Internet, subsequently takes part in the local reconstruction activities. Meanwhile there are nodes' and motifs' death. In process of the local reconstruction, although the Internet system replicates motifs repeatedly by adding or removing actions, the system characteristics and global structure are not destroyed. Statistics about the motif *M3* which is a full connectivity subgraph shows that the process of its metabolism is fluctuation that causes mutation of Internet. Furthermore we find that mutation is instinctive reaction of Internet when it's influenced from inside or outside environment, such as Internet bubble, social network rising and finance crisis. The behaviors of metabolism, self-replication and mutation of Internet indicate its life characteristics as a complex artificial life. And our work will inspire people to study the life-like phenomena of other complex systems from angle of topology structure.

## 1 Introduction

Life is one beautiful existence and the research about it is a never ending subject for the biologist, anthropologist, medical scientist, philosopher, physicist and many others. One of the important areas of research is, what signs it should have, if we call it living things. Researchers from different research areas want to answer this question with general law. In fact people have gained significant achievements to explain the origin of life such as Darwin's evolution theory[1][2], Hyper-cycle theory[3], self-organization[4] etc. . Although they describes the living process from different angle, the common is to explore the origin and evolution of living things.

The nature life, such as human, animal and microorganism, is the earliest research object in the life sciences. In 19th century the cell theory was proposed by Matthias Jakob Schleiden and Theodor Schwann which describes the properties of cells. Three tents to the cell theory are described: (1) all living organisms are composed of one or more cells; (2) cell is the basic unit of structure and organization; (3) cells come to existence from pre-existing cell[5]. They reveal the metabolism and self-replication phenomena of living organisms from the cell level. Subsequently researchers discovered the principle of cell growth that is parent DNA replicates themselves to child DNA, then the DNA and protein are put together to form a new cell[6]. In the process of DNA replication, mutation may occur because of the genetic changes[7]. Meanwhile in the study of protein components scientist discovered protein motif structure that, there are many similar chain-like biological molecules in protein which describes the connectivity between secondary structural elements[8].

With the development of modern physics, more and more people find that some non-life systems present the life-like behaviors such as dissipative structure[9], synergetic[10],mutation[11], emergence[12], and self-organization[4]. They are called complex systems. The hyper-cycle theory proposed by M. Eigen described the life process of the non-life complex systems from the angle of biochemistry, which includes three cycles and mutation process. From the modern views, if one complex system has the behaviors of metabolism, self-replication and mutation, it is a life body. It also points to the system's evolution under competition and environment selection.

In the past decade network science has become an important means to research complex systems[13][14][15]. The network topology or model abstracted by node entity and existing relationship are main research object. It's amazing that Shen-Orr et.al[16] introduced structure motif into complex networks—network motif—and discovered the network motifs of gene regular network. In 1999, Faloutsos et al. [17] found the power-law of degree distribution of Internet topology and refuted people's old view that the Internet is a random network. Now the Internet has become such a giant complex system [18] that nobody can describes it. However researchers keen on exploring the dynamics of Internet evolution from topology analysis, modeling, Internet measurement and economic ecosystem. No one studies the life-like phenomena of Internet as one of the largest man-made complex system. In this paper, we attempt to research the life characteristics of Internet, as artificial life, by network motif and the hyper-cycle theory. In section 2, we review the related work about the life characteristics, motif and Internet evolution. Section 3 proposes a framework of complex network evolution that describes a novel method, which is appearance or disappearance of node, edge and local structure, to research network evolution on time-series. Based on the above framework, Internet metabolism, self-replication and mutation behaviors are analyzed by node, edge and motif. In the section 4. Section 5 provides some concluding remarks and future work.

## 2  Related work

The earliest precursor to modern complex systems theory can be found in the classical political economy of the Scottish Enlightenment. Now such systems are used to model processes in computer science, biology, economics, physics, chemistry, and many other fields. In late of 1960s, L.Von Bertalanfy proposed the self-organization theory which studies the complex phenomena of nature and society and explores the basic law of complex phenomena during forming and evolution. This theory is composed of dissipative structure[9], synergetic[10], mutation[11] and hyper-cycle[3]. Especially the hyper-cycle theory describes the evolution process from non-living to living system.

Although different theories focus on different parts of complex system, all of them want to discover the principle of complex system, i.e. how to organize, evolve and manager themselves whether it is influenced by other factors or not. The Ref. [17] firstly proved Internet is a scale-free network topology in 1998. K. Park et.al [18] researched the Internet as a complex network from the aspects of self-similar traffic, power-law connectivity, WLAN PHY layer dynamics and non-cooperative network games. However most researchers study the dynamics of Internet evolution from two angles: physical network layer and logical topology layer. The research about physical network mainly focuses on the Internet economical ecosystem, especially the AS relationship. Gao[19] proposed the AS relationship inference algorithm firstly. Then Ref. [20] analyzed the Internet hierarchy structure according to put customer-provider and peer-peer into five-level classification of ASes. And Ref. [21] proposed a network aware, macroscopic model that captures the characteristics and interactions of the application and network providers which lets us know the cause and effect of the Internet evolution. A. Dhamdhere et.al analyzed the AS-level Internet growth for over a decade based on AS function and business type [22][23]. However the others research Internet based on complex network theory, for example the Ref. [24][25] verify the scale-free of preferential attachment of complex network by analyzing the Internet evolution. However the above research doesn't consider the systematic behavior of Internet. In the Ref. [26], the author had researched phase changes phenomena of IPv4 and IPv6 during the process of its growth. And the Ref [27] discovered the sudden change of IPv6 topology based on mean degree, mean path length and other metrics on time-series. Both of them only research the fluctuation or mutation of Internet as a complex network.

At the same time, Moore et.al [28] proposed a more realistic model—time evolution model of networks—by addition and deletion of nodes. Here people started to realize that the essence of network evolution is the changes of nodes and edges in the topology. In Ref. [29], the authors formulated the topology liveness problem and contrasted it with the completeness problem. And they developed an empirical model that characterize the changes in observed AS topology by three process: birth, death and revelation. As we all know, only if one born node builds more connections with others, it could be more robust. It is just like that a man would like to build new interpersonal relationship by making friend when he enters into the strange environment. So some people research fundamental function structure by analyzing the subgraphs or components of complex network[30][31][32]. And network motif, one type of subgraphs, is better to present the particular pattern of network. Currently the researchers are high on the motif detecting algorithm and motif analysis of network, for example the protein motif[33], gene motif[34], software motif[35], social motif[36] etc. . In this Ref. [37], topological patterns were used to detect network feature and identify hierarchical feature of Internet topology. And R. Kiremire et.al[38] adapted the network motif approach to provide a common ground for comparing the schemes' performances in different Internet topologies.

## 3   Framework of Complex Network Evolution

For a complex network $g(i)$, $i$ denotes one number of network snapshot which is extracted at time $t$, $N(i)$ is the node set of network $g(i)$ and $E(i)$ is the edge set of network $g(i)$. So a series of networks $G = \{g(1), g(2), g(3), \ldots, g(n)\}$ which are ordered on the time are obtained by extracting the network snapshot on fixed time window $\nabla t$. As a complex system, it is always growing to satisfy the human needs. And the basic and obvious is nodes' and edges' change to drive network evolution. The Fig. 1 is the schematic diagram of network evolution from $T(i)$ to $T(i+1)$, where $T(i+1) - T(i) = \nabla t$. From snapshot $g(i)$ to $g(i+1)$, the network goes through a dynamic change. Then in this process, if some nodes of network are steady, we call them **steady-nodes** $N_{steady}(i, i+1) = N(i) \cap N(i+1)$, such as {*e, h, g, f, i*}. If some nodes in the network disappear at time $T(i+1)$, they are called **dead-nodes** $N_{dead}(i, i+1) = N(i) - N(i+1)$ such as {*a, b, c, d*}. The **born-nodes** $N_{born}(i, i+1) = N(i+1) - N(i)$ are nodes which appear at time $T(i+1)$, such as {*j, k, l, m*}.

The nodes' birth and death must result in some edges' change, such as the direct edges which are those connections between born or dead node and network structure. Due to the two endpoints of one edge belong to different node sets possibly, the edges are classified into three categories: **outer-edge**, **boundary-edge** and **inner-edge**. **Outer edges** are the connections which both of endpoints are in

$N_{dead}(i,i+1)$ or $N_{born}(i,i+1)$, it suggests that the local subnetwork as a function structure maybe added or removed. The **boundary edges** disappear or appear in next snapshot because of the existed steady nodes. The **inner edges** are the connections in which two endpoints of one edge are steady node, but the connection relationship changes. So there are six type edges that are extracted from network $g$, i.e. $E_{death}^{outer}$, $E_{death}^{boundary}$, $E_{death}^{inner}$, $E_{birth}^{outer}$, $E_{birth}^{boundary}$ and $E_{birth}^{inner}$:

$$\begin{cases} E_{death}^{outer}(i,i+1) = \{e \in E(i) \mid n_e^l \in N_{death} \text{ and } n_e^r \in N_{death}\} \\ E_{death}^{boundary}(i,i+1) = \{e \in E(i) \mid n_e^l \in N_{death} \text{ and } n_e^r \in N_{steady}\} \\ E_{death}^{inner}(i,i+1) = \{e \in (E(i) - E(i+1)) \mid n_e^l \in N_{steady} \text{ and } n_e^r \in N_{steady}\} \\ E_{birth}^{outer}(i,i+1) = \{e \in E(i+1) \mid n_e^l \in N_{birth} \text{ and } n_e^r \in N_{birth}\} \\ E_{birth}^{boundary}(i,i+1) = \{e \in E(i+1) \mid n_e^l \in N_{birth} \text{ and } n_e^r \in N_{steady}\} \\ E_{birth}^{inner}(i,i+1) = \{e \in (E(i+1) - E(i)) \mid n_e^l \in N_{steady} \text{ and } n_e^r \in N_{steady}\} \end{cases} \quad (1)$$

Where $n_e^l$ and $n_e^r$ is left and right endpoint of edge $e$. Meanwhile parts of edges not only build the motif structure, but also form a complex subnetwork. For example the inner edges form inner-subnet.

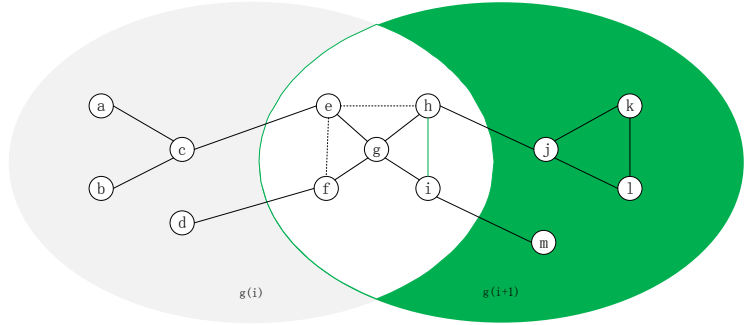

Fig. 1 Schematic diagram of network evolution on the time.

Ref. [28][29] only researched the birth and death of nodes. According to Table 1, we find that not only the edges are important components of network evolution, but also the local subgraph structure which is composed of several nodes have a significant influence for network growth. Motif has been suggested to the functional building block of network complexity. Network motif, patterns of interconnection occurring in complex networks at numbers that are significantly higher than those in randomized networks, are defined to uncover structural design principles of networks. In this paper, we research the three-node motif and one pattern with two node shown in Fig. 2. For a general terminal device *a*, it only need to connect the device *b* to access Internet. One router or network server, as the Fig. 2(b) shows, would build multi-connections with other devices. However some key routers, especially the Tier-1 routers, may structure the triangle for improving the Internet robustness and decreasing the cost of data transmission.

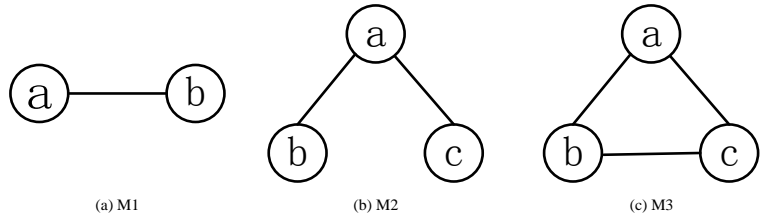

Fig. 2 Two-node pattern and Motif with three nodes in undirected network

## 4 Life Characteristics of Internet

From four experimental nodes in 1969 to 1983's three hundred connected computers until enough Internet terminals currently, Internet has become a giant and complex system consisting of tens of thousands of Autonomous System (AS) and billions of IP address. Nowadays Internet changes dynamically according to the humans' need, and nobody can describe this system. So it has very significant value for researching life characteristics of the complex Internet.

Our work focuses on the AS-level Internet topology because AS-level dataset is the most enough. The topology is extracted from the BGP router table data which is collected by Route-Views[39], RIPE NCC[40]and PCH[41] project. Firstly all probing data from January 1988 to August 2013 are downloaded. Then we combine data in a month as one snapshot of Internet. Finally according to extracting the AS_PATH field and filtering the poisoned data[22], 184 snapshot topologies are obtained which are ordered on time-series. The following work is done based on the framework of complex network evolution proposed at section 2.

### 4.1 Metabolism

In biology, Metabolism is a term that is used to describe all chemical reactions involved in maintaining the living state of the cells and the organism. It can be conveniently divided into two categories: Catabolism (breakdown of molecules to obtain energy) and Anabolism (synthesis of all compounds needs by the cells). For one growing complex network gained from complex system, it must exist the topology structure that transmits signal, information or matter from source to target node effectively and quickly, such as communication network, social network, power grid, protein-protein interaction network, neural network etc. So nodes and edges are the fundamental elements of complex network. In general, one node is a producer, transmitter or both and the edge is the bridge of information transmission. Then network topology is a system what is always dispatching the path of forwarding information flow. Therefore the global topology can influence the local flow direction and node load, and the local changes give rise to the dynamic of system. For example the social needs for Internet promote the ISP to build more network in the local area because the Internet have improved the production efficiency of human society and promoted the economic development quickly. The more abundant the Internet infrastructure is, the more faster the economy grows [42]. Due to the router can't connect to infinite routers or the old architecture can't adapt to modern technology, the ISPs will restructure the local network or change the Internet architecture to decrease the complexity of network when it grows to a certain stage[43]. Here, we simplify above complex process into following model: addition of the nodes, edges and motifs break the equilibrium state of system; removing some nodes, edges and motifs is to maintain the system's equilibrium. We call this process **Internet Metabolism**.

Based on above analysis, we know that the direct representation of Internet metabolism is birth and death of node, edge and motif. The Fig. 3 shows the Internet metabolism process from January 1998 to August 2013. We find, as the Fig. 3(a) shown, the Internet is always growing dynamically and the number of born node is bigger than that of dead node. The born node has same curve with the dead node on time-series. This curve indicates that Internet grows regularly and steadily to maintain system stability.

The Fig. 3(c)-(d) is the edges' metabolism including inner-, boundary- and outer-edge. The inner-edge is that mainly takes part in the local restructure to make the Internet not only meet the requirement of system growth but also improve system robust. The boundary-edge ensures one node join or leave the Internet system. The outer-edge explains that the Internet metabolizes with some complex structures, not just the discrete nodes. The rate of three types of edges show that the $E^{inner}$, about 90%, is the most active component in process of Internet metabolism, then the about 9% of $E^{boundary}$. The least is $E_{born}^{outer}$ and $E_{dead}^{outer}$ depicted in the Fig. 3(d). It implies that the inner-edges dominate the Internet metabolism from simple to complex beside node birth and death. We find the trend of metabolism of $E_{dead}^{inner}$ and $E_{born}^{inner}$ can be depicted with exponential equation $y = ae^{bx} + c$, i.e. $y_{born} = 712.34 * e^{x/79.97} + 253.41$ and $y_{dead} = 1064.67 * e^{x/103.65} - 390.44$. Then the mean birth rate from 1998 to 2013 is calculated by equation(2) $r = 0.96\%$. It indicates that the Internet system is in an equilibrium state of structure evolution, although the number of node increases exponentially. The trends of Fig. 3(c) is similar to Fig. 3(a) because one born node have to build connections with the existing nodes and dead nodes cut off the connections from Internet system.

$$r = \frac{1}{n-1} \sum_{i=1}^{n-1} \frac{|E_{born}(i,i+1) - E_{dead}(i,i+1)|}{|E(i)|} \quad (2)$$

According to researching Internet metabolism in the aspect of node and edge evolution, we find that one born node connect one or more existing nodes to be a part of Internet system firstly. Then it takes part in the local restructure activity that build more connections by way of motif. Although the nodes and edges are the basic elements of complex network, the motif which represents the function structure is more significant for researching the life characteristic of Internet.

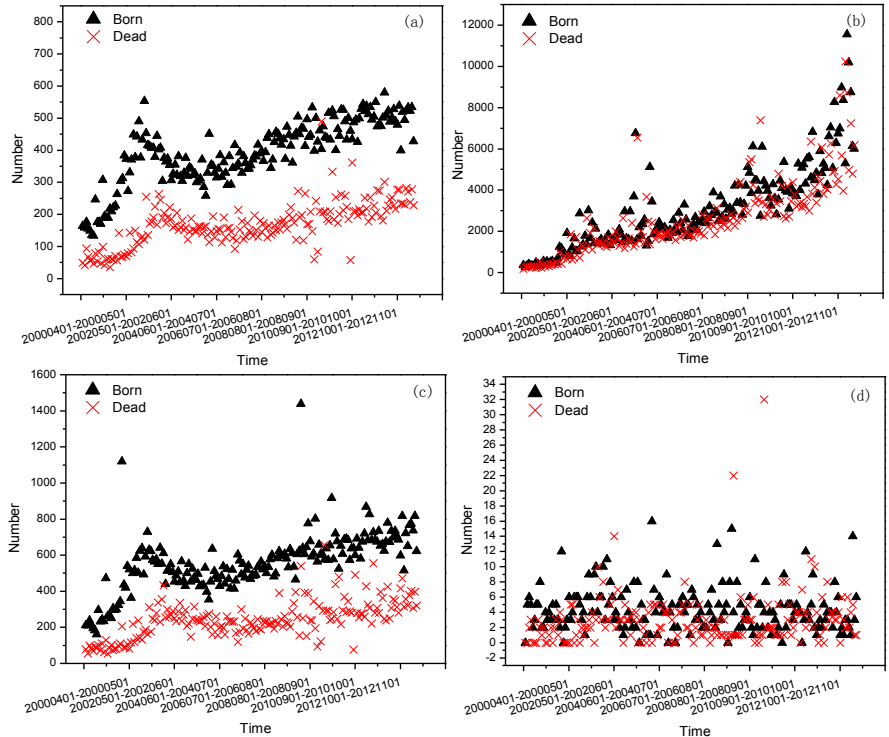

Fig. 3 Metabolism of nodes and edges of network. (a) network nodes, (b) network inner-edge,
(c) network boundary-edge, (d) network outer-edge.

In this paper, we research the pattern with two nodes (M1) and motif with three nodes (M2 and M3) as the Fig. 2 shows. Here the motifs of boundary-subnet are different from others. For the M1, it must be one endpoint belonging to $N_{steady}$ and the other belonging to $N_{born}$ or $N_{dead}$. And one node of the motif M2 belongs to $N_{steady}$ and the others belong to $N_{born}$ or $N_{dead}$. The Table 1 analyzes the metabolism of M1, M2 and M3 by 16 snapshots of Internet topology. The Fig. 3(d) shows that the number of outer-edge is much less than that of the inner- and boundary-edge, so the number of outer-motif is also very few. Due to our definition in section 3, the motif M3 doesn't exist in boundary-subnet. However the number of born M2 is far greater than that of born M1 and the number of dead M2 is similar to that of dead M1. It suggests that one born node prefers to build connections with more than two nodes and the dead nodes are the leaf nodes of network. As we all know, the more connections one node has, the more robust it is.

Table 1 Metabolism of M1, M2 and M3.

| | Birth | | | | | | | | | Death | | | | | | | | |
|---|---|---|---|---|---|---|---|---|---|---|---|---|---|---|---|---|---|---|
| | Inner | | | Boundary | | | Outer | | | Inner | | | Boundary | | | Outer | | |
| | M1 | M2 | M3 | M1 | M2 | M3 | M1 | M2 | M3 | M1 | M2 | M3 | M1 | M2 | M3 | M1 | M2 | M3 |
| 19980201-19980301 | 268 | 1240 | 0 | 202 | 1272 | 0 | 10 | 0 | 0 | 253 | 860 | 0 | 80 | 98 | 0 | 0 | 0 | 0 |
| 19990201-19990301 | 423 | 1980 | 6 | 251 | 1636 | 0 | 6 | 0 | 0 | 343 | 1238 | 0 | 57 | 112 | 0 | 0 | 0 | 0 |
| 20000201-20000301 | 733 | 34952 | 306 | 350 | 3658 | 0 | 4 | 0 | 0 | 561 | 15286 | 156 | 117 | 168 | 0 | 0 | 0 | 0 |
| 20010201-20010301 | 1170 | 64744 | 846 | 597 | 19766 | 0 | 12 | 0 | 0 | 921 | 14122 | 132 | 166 | 642 | 0 | 4 | 0 | 0 |
| 20020201-20020301 | 1198 | 20794 | 108 | 440 | 6260 | 0 | 10 | 0 | 0 | 1038 | 18172 | 108 | 229 | 1546 | 0 | 6 | 0 | 0 |
| 20030201-20030301 | 1286 | 36086 | 240 | 379 | 3748 | 0 | 6 | 0 | 0 | 1127 | 76520 | 900 | 180 | 470 | 0 | 2 | 0 | 0 |
| 20040201-20040301 | 1826 | 531222 | 29328 | 559 | 5652 | 0 | 23 | 20 | 0 | 1088 | 53330 | 222 | 170 | 476 | 0 | 0 | 0 | 0 |
| 20050201-20050301 | 1380 | 59250 | 330 | 396 | 1902 | 0 | 4 | 0 | 0 | 1363 | 36546 | 330 | 255 | 1008 | 0 | 0 | 0 | 0 |
| 20060201-20060301 | 1447 | 113216 | 1458 | 464 | 2298 | 0 | 6 | 0 | 0 | 1336 | 32308 | 192 | 211 | 468 | 0 | 4 | 0 | 0 |
| 20070201-20070301 | 1836 | 94338 | 990 | 504 | 4570 | 0 | 7 | 22 | 0 | 1614 | 22274 | 102 | 257 | 412 | 0 | 2 | 0 | 0 |
| 20080201-20080301 | 2226 | 123778 | 1794 | 649 | 2338 | 0 | 4 | 0 | 0 | 1632 | 94996 | 1362 | 183 | 354 | 0 | 2 | 0 | 0 |
| 20090201-20090301 | 2248 | 303998 | 2742 | 616 | 2628 | 0 | 8 | 2 | 0 | 1752 | 226644 | 3462 | 158 | 252 | 0 | 2 | 0 | 0 |
| 20100201-20100301 | 2023 | 510972 | 9090 | 631 | 2572 | 0 | 0 | 0 | 0 | 1897 | 352470 | 3828 | 339 | 1238 | 0 | 13 | 6 | 0 |
| 20110201-20110301 | 2297 | 585482 | 6456 | 764 | 2732 | 0 | 3 | 6 | 0 | 2200 | 88306 | 2346 | 331 | 482 | 0 | 6 | 2 | 0 |
| 20120201-20120301 | 2582 | 340956 | 4266 | 746 | 1978 | 0 | 6 | 0 | 0 | 2323 | 215892 | 672 | 295 | 608 | 0 | 2 | 2 | 0 |
| 20130201-20130301 | 2731 | 392744 | 3492 | 742 | 1930 | 0 | 6 | 0 | 0 | 2425 | 2755530 | 59280 | 391 | 912 | 0 | 7 | 20 | 0 |

For the inner-subnet, the motif M2 is the main metabolism structure that accounts for 95%~97% and the M3 is least (1.04%). These results indicate that motif M2 is the fundamental function unit in process of Internet metabolism. This also confirm a fact that star- or tree-structures are most common in the Internet, i.e. one node prefers to connect multi-nodes to form motif structure in process of taking part in the network restructure, then one node of motif connect to the other node which is in the higher level.

## 4.2 Self-replication

Self-replication is a behavior of dynamic system that construct an identical copy of itself. Biological cells, given suitable environments, reproduce by cell division. During cell division, the DNA structure is copied and can be transmitted to offspring during reproducing. Biological viruses can replicate, but only by commandeering the reproductive machinery of cells through a process of infection. Harmful prion proteins can replicate by converting normal proteins into rogue forms. Computer viruses reproduce using the hardware and software already present on computers. Does the Internet shows the behavior of self-replication as complex system? If it does, what is the process of replicating network structure?

In section 4.1, the Internet metabolism behavior is analyzed from node, edge to motif structure based on the Internet evolution. It not only receives the new node to join system, but also build some motifs to reconstruct system structure. Table 1 shows that the Internet construct motif repeatedly to meet the requirement of system changes. For example, the M2 can form a simple star, or complex tree. And M3 is a triangle which is also stable structure. The more M3 the Internet produces, the more robust it become[44]. So adding node and constructing motif are main factors for network anabolism. We call them **Internet self-replication**.

How does the Internet replicate itself? To start with this problem, the evolution of a node is analyzed from the first snapshot to the last. One new node is like a newborn who grows and makes friends with others, encounters some abnormal events and is restrained by the surrounding environment. So some nodes may disappear after a short period of time. But most nodes take part in the system metabolism continuously, even some of them become the important nodes of network. In the Fig. 4, the AS7713[1] is added into Internet at Journal 1998, then grows exponentially after October 2004. It suggests that a born node adapts the system environment for a while, then grows quickly to become the important node. Fig. 4(c) is a visualization of local network including the AS7713 and its neighbors in the snapshot network at August 2013.

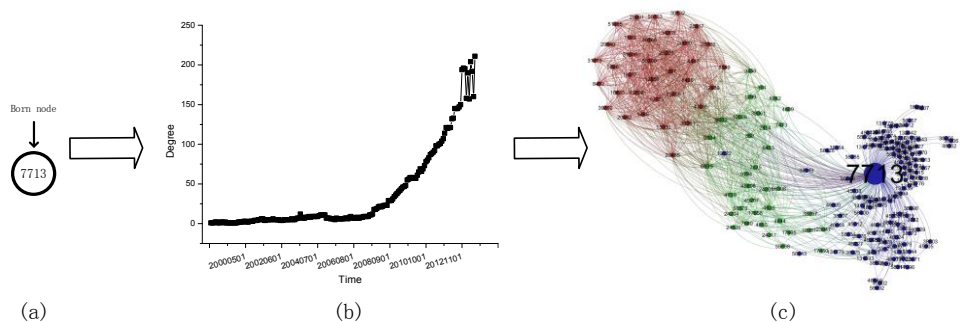

Fig. 4　Growth process of AS7713.

Next, the network reconstitution is analyzed when a born node wants to take part in more evolution activities. Table 1 indicates that the inner-edges are the fundamental part of Internet metabolism, and the boundary-edges are used to receive the born nodes or remove the dead nodes. The Fig. 4(b) shows that AS7713 become a central node after a long time, no but immediate. So the local network with AS7713 should restructure themselves by adding or removing motif structures for a long time. Thus we extract the some reconstitution processes in the AS7713 local network.

Fig. 5 is three cases of inner-edge reconstitution in AS7713 local network from 2010 to 2012, where the edges are divided into three types: steady edges, born edges and dead edges. In Fig. 5(a), the motif M2 ({AS6939-AS38202, AS38202-AS4844}, {AS6939-AS38202, AS38202-AS4826} and {AS6939-AS38202, AS38202-AS4826}) are removed from Internet, meanwhile the motif M3 ({AS55658-AS4844，AS4844-AS4826，AS4826-AS55658}) is added. This let AS7713 access the AS4844 and AS4826 via AS55658, although the connections of AS38202-AS4844 and AS38202-AS4826 are cut off. The local network changes from the motif ({AS23947-AS9304, AS23947-AS45896}) to the motif ({AS23947-

---

[1] AS7713 represents the autonomous system number of a single administrative entity or domain whose information is registered in the RIR.

AS55658, AS23947-AS18351}) in Fig. 5(b). However for Fig. 5(d), the AS4844-AS2411 is deleted and the M1 and M2 are added. The first one improves the network robustness because of the stability of M3 structure, the second one changes the connection and the third enhances the network connectivity by introducing more inner edges.

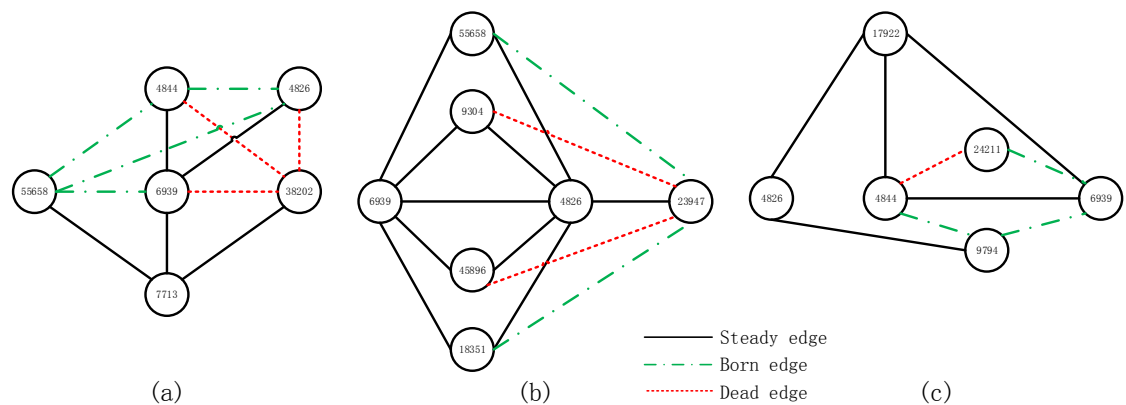

Fig. 5 Diagram of local network (AS7713) reconstruction.

Self-replication is to replicate the M1, M2, M3 structures repeatedly and maintain the system characteristic. So the problem, whether the Internet self-replication behavior changes the Internet system characteristic or not, need to be validated from the angle of global topology structure. The Internet is a representative complex network with scale-free characteristic [17]. Its most important feature is that, the degree distribution of the scale-free network is power-law. Here the power-law of Internet is in [1.78, 2.25] by calculating all snapshots topology, and the standard deviation is 0.13. Results show that Internet topology structure is stable globally.

## 4.3 Mutation

In biology, a mutation is a permanent alteration of the nucleotide sequence of the genome of an organism, virus, extrachromosomal DNA or other genetic elements. Mutations may or may not produce discernible changes in the observable characteristics of an organism. Mutations play a part in both normal and abnormal biological processes including: evolution, cancer, and the development of the immune system, including junctional diversity. The Mutation theory is proposed from *structure stability and morphogenesis* written by R. Thom at 1972. R. Thom thinks that mutation is one jump of inter-state where the system is considered overall. Its characteristic is continuous process and uncontained result of internal system[11]. Then Mutation concept is also introduced into hyper-cycle theory by M. Eigen at 1977. The main source of mutations, especially in the early stages of evolution, is miscopying, i.e. the inclusion of a nucleotide with a non-complementary base during the process of replication. So mutation is a sudden change of system in process of self-replication which brings the system into stable equilibrium from unstable state. For the Internet topology, its metabolism and self-replication behaviors must occur with Internet reconstruction activities at every moment. As complex system, Internet maintains the system stability regardless of whatever reconstruction is going on. So it has a critical state from quantitative to qualitative.

As we all know, triangle is the most stable structure in our world. If there were existing triangles in every three nodes, the Internet topology would be most robust. Due to the geography, policy and economy, this scheme is unrealistic. So we still think that the motif M3 is the most important components of Internet, although its quantity is lower than 2%.

In Fig. 6, the motif M3 evolution is analyzed from January 1998 to August 2013. The most obvious is that it will have a larger rate of birth subsequently, if the death of M3 happens abnormally, and vice versa. If M3's birth is anabolism, then its death is catabolism. In this paper Internet mutation occurs, if the rate of birth or death is greater than 3% during once analysis. But the rate of birth and death is mostly in the equilibrium state. For the anabolism mutation, Internet becomes more and more steady, the catabolism mutation makes Internet become more orderly. Here the structure entropy is used to evaluate this sudden change. For two mutations marked in the Fig. 6, the rate of M3 death is 4.99% in the June 2005's snapshot topology(M3 birth is 0.70%), then rate of M3 birth is 6.71% in next snapshot(M3 death is 1.62%). Respectively the normalization structure entropy is from 0.1756 to 0.1654. It implies that adding the M3 improves the system stability.

Furthermore, we research the relationship between Internet mutation and Internet economy. All the mutations are classified into three time period: 2000~2003, 2005~2006 and 2008 shown in the Fig. 6. At 2000~2003, the Internet bubble was a historic speculative bubble during which stock markets in industrialized nations saw their equity value rise rapidly from growth in the Internet sectors and related fields. At the beginning, the ISPs expand the Internet infrastructure abnormally to meet the needs of Internet Company. After the Internet bubble, Internet resources were excess seriously because of a large number of Internet companies going broke. For decreasing the operation cost and improving the efficiency of network transmission, some ISP companies chose to merge each other and others optimize self-network topology and provide better network services. Whatever the merger or optimization was done, the Internet topology had changed to response the effects from surrounding environment. Fig. 6 shows that there are multi mutation including four death events and three birth events. The 2006 year is as the Time magazine's Person of the Year because the social network is becoming the more and more popular, such as Facebook, Twitter, Myspace and YouTube. But the 2008 financial crisis brought Internet into the new development of low tide.

In summary, the Internet as part of earth ecosystem have to change self-structure by mutation behavior to ensure its best operation status, when it is inferenced by outside environment.

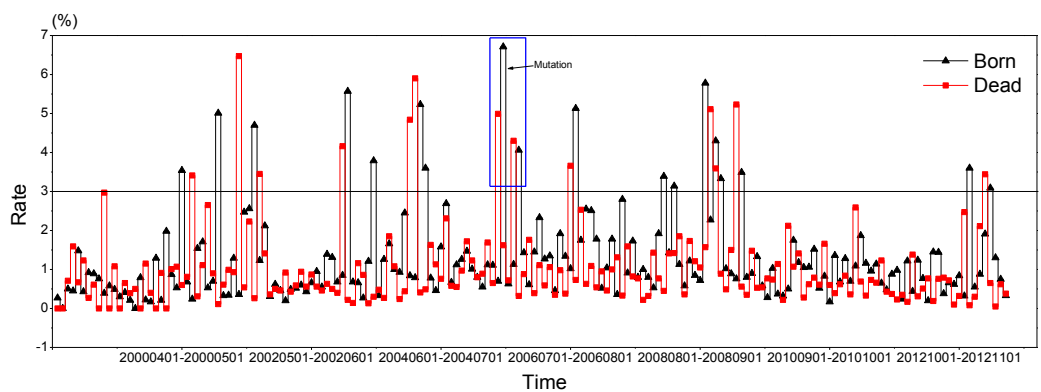

Fig. 6 Internet mutation in process of motif M3 restructure

## 5 Discussion and Conclusion

The complex networks abstracted from complex systems have been an important research direction. In this paper, we researched the life characteristics of non-life, complex system from the angle of system topology. Due to lack of dynamic network topology dataset, especially insufficient long term data, practice work would be very hard. Here we extract 184 snapshots of Internet from January 1998 to August 2013. These rich data make it possible to research the life characteristic of Internet.

Based on the hyper-cycle theory, the non-life complex system is living if there are the metabolism, self-replication and mutation characteristics. To research these characters, firstly a framework of evolution analysis of complex networks is proposed based on set theory and graph theory. It provides a general and novel method for network evolution research because of considering the node, edge and motif of network topology.

Birth and death of network node, edge and motif structure is external manifestations of Internet metabolism. In our analysis, node metabolism activities mainly located at the border of Internet system, i.e. born and dead nodes are some those with small degree. It infers that one new node only join the Internet system at the first step. The Table 1 shows that 95% of motif metabolism happen within the Internet system and the others participate in node metabolism activities. The result indicates that there are two types of structure metabolism: metabolizing with node and without node. It also means that most new nodes participate in the network construction after they are added. It is the first step for a new added node, and the final goal is network restructure base on function structure—motif. Here we call the behavior of copying motif structure repeatedly as Internet self-replication. The Fig. 5 shows the restructuring activities of motif in the local network of AS7713 during process of Internet metabolism. The M2 builds the star and tree topology and the M3 improves the system stability because of the rectangle principle. The Internet self-replication behavior doesn't change the global topology, but restructure local path of information transmission. The power-law of degree distribution calculated verifies that self-replication is a behavior of maintaining system function in the Internet metabolism. But being affected from external environment inevitably, the Internet also happens system mutation correspondingly, just as most ecosystems. Here, the rate of birth and death of M3 is used to evaluate the Internet mutation. And the M3's mutation obviously corresponds the three important social events: Internet bubble in 2000~2003, social network rising in 2005~2006, financial crisis in 2008.

According to long term analysis of metabolism, self-replication and mutation behaviors, we find that the Internet possesses life characteristics as a non-life complex system. For the first research done from angle of topology structure of complex system, we will continue the research work about Internet life characteristics. The M1, M2 and M3 structure would be the most fundamental function component of complex structure, however our future work will focus on the complex motifs, such as motifs with four, five and six nodes.